\documentclass{networks2007}
\usepackage{amssymb}
\usepackage{graphicx}

\begin{document}


\newcommand{\Eqref}[1]{(\ref{#1})}


\markboth
  {Template for the Net-Works 2007 proceedings} 
  {M. A. Dahlem, F. M. Schneider, A. Panchuk, G. Hiller, E. Sch\"oll}                


\title{Control of sub-excitable waves in neural networks
by nonlocal coupling}



\author{Markus~A.~Dahlem}{dahlem@physik.tu-berlin.de}{{1,2}}

\author{Felix~M.~Schneider}{schneider@itp.physik.tu-berlin.de}{1}

\author{Anastasiya~Panchuk}{nastyap@imath.kiev.ua}{{1,3}}

\author{Gerald~Hiller}{hiller@itp.tu-berlin.de}{1}

\author{Eckehard~Sch\"oll}{schoell@physik.tu-berlin.de}{1}


\affiliation{1}{Institut f\"ur Theoretische
              Physik}{Technische Universit\"at Berlin, Germany}

\affiliation{2}{Klinik f\"ur Neurologie II}{Otto-von-Guericke-Universit\"at Magdeburg, Germany}

\affiliation{3}{Institute of Mathematics}
{National Academy of Sciences of Ukraine}


\begin{abstract}

Transient wave forms in neural networks with diffusive and nonlocal
coupling have attracted particular interest because they may mediate
recruitment of healthy cortical tissue into a pathological state
during migraine.  To investigate this process, we use a
reaction-diffusion system of inhibitor-activator type as a generic
model of pathological wave propagation and set it close to bifurcation
in the sub-excitable regime. We report the influence of various
nonlocal connectivity schemes on wave propagation.  Wave propagation
can be suppressed with cross coupling inhibitor and activator for both
positive and negative coupling strength $K$, depending on the
connection length $\delta$. The area in the parameter plane
$(\delta,K)$ where this control goal is achieved resembles a
Mexican-hat-type network connectivity.  Our results suggest that
nonlocal synaptic transmission can control wave propagation, which may
be of therapeutic value.

\keywords nonlinear  dynamical systems, excitability,  control,
nonlocal and time delay coupling
\end{abstract}

%
%

\section{Introduction}

During migraine attacks, localized pathological excitation can spread
through cortical tissue and invade large areas before it abates.  This
activity causes migraine aura, that is, neurological symptoms
preceding the headache phase \cite{LAU94}.  The underlying process is
a phenomenon called cortical spreading depression (SD).  It is assumed
to be a reaction-diffusion process in the cortex, although reactions
and diffusion processes that provide the mechanism of propagtion are
still under debate \cite{STR05,HER05}. However, the generic dynamics
of reaction-diffusion systems are largely independent of the
interaction details and shared among various biological systems
\cite{HES00}.  Therefore, to describe the spatio-temporal patterns of
SD, the cortex can be approximated as a continuous excitable media
supporting reaction diffusion waves \cite{REG94,DAH03a,DAH04b}.

Psychophysical studies on visual processing in migraine patients
suggest that changes in their networks of cortical neurons lead to an
interictal state of changed excitability, i.\,e., an anomalous
cortical state in the interval between migraine attacks
\cite{WEL90,SHE01}.  This motivates efforts to understand how the
spread of reaction diffusion waves is controlled by nonlocal network
connectivity. To include this, we investigate in this work the
hypothesis that the emergence of SD waves can be attributed to these
network changes as well.  Previously, we have investigated how to
change parameters of an excitable medium so as to efficiently protect
cortical tissue surrounding a stimulus against recruitment
\cite{DAH07a}.  Our current results suggest that failures in synaptic
transmission result in increased susceptibility of cortical tissue to
SD. Such a modulation of excitability becomes of crucial importance
when the cortical state is close to the bifurcation of the onset of
wave propagation. The clinically relevant conclusion to be drawn from
this is that therapy might target network connectivity that modulates
cortical tissue excitability, even though a specific network
connectivity is not required for the initiation or propagation of SD in a
regime far from the bifurcation.

\begin{figure}[!tpb]
\centerline{\includegraphics[width=\textwidth]{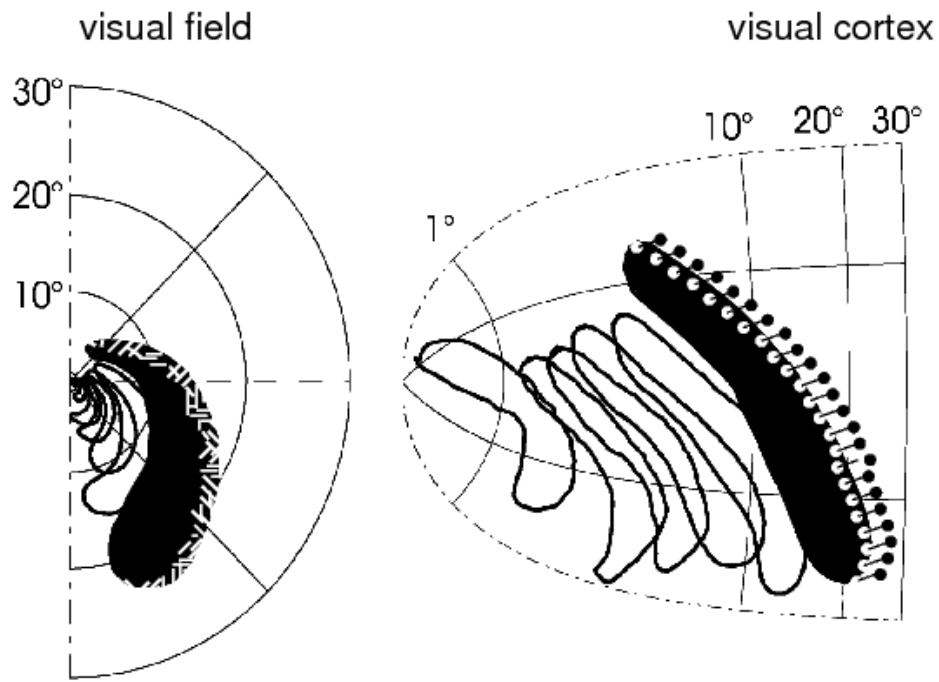}}
\caption{\label{fig:ret} (left) Visual migraine aura symptom in the
shape of a jagged crescent pattern moving through the visual field
(modified from \cite{LAS41}). (right) Translation of the visual field
disturbance by inverse retinotopic mapping. The crescent patterns
translates into a wave segment resembling a ``critical finger''
\cite{HAK99}. Such patterns are unbounded wave segments observed close
to the excitability boundary, i.\,e., a bifurcation of the systems
described by Eqs.~(\ref{eq:fhnPlusNLC-1}-\ref{eq:fhnPlusNLC-2}).  }
\end{figure}

\section{Neural network with diffusive and nonlocal
coupling}

A variety of neural network models for SD have been proposed, though
there is not yet consent on the mechanism. Roughly speaking, two
classes of models exist.  One taking a bottom up approach based on
biophysical laws including several ionic currents, ion pumps, membrane
potentials, and osmotic forces \cite{Tuc78,Kag00,Som01,SHA01}. The
other approach is top down trying to incorporate the system properties
without detailing any first-level subsystems. Hodgkin proposed the
first simplified reaction-diffusion approach to SD. The model was
never published, but communicated to Grafstein \cite{GRA63,BUR74}.
According to this references, Hodgkin suggested to consider a balance
equation for potassium with a cubic source function and diffusion.
The three roots of the cubic function being the resting state, a
threshold, and a potassium ceiling level, respectively.  Based on
methods introduced by Huxley, solving the equation led to the
approximate speed of SD.  However, the diffusion coefficient used was
four times higher than that of potassium in aqueous solution. This may
either indicate the anomalous nature of potassium migration in the
cortex \cite{NIC81}, or the leading role of a faster transcelluar
reaction-diffusion mechanism \cite{HER94}.  

Another model \cite{REG94} followed Grafstein's potassium hypothesis
with two extensions.  We shortly summarize this work, because we
follow a similar approach but with a different interpretation and aim.
Firstly, the model included a second dynamic variable describing a
refractory phase of SD.  The source term in the balance equation of
potassium is replaced by a quartic polynomial as the major
nonlinearity of this activator-inhibitor system.  Secondly, the
reaction-diffusion model was connected to a neural network building
together a hybrid model. The neural network was originally used to
study cortical dynamics and sensory map reorganization.  In the hybrid
model it was used to explain visual field defects occurring during
migraine with aura.  We also use a reaction-diffusion system combined
with a nonlocal interaction coming from a neural network.  Our goal is
to investigate which neural network connectivity can prevent SD, as
suggested and studied using cellular automatons in \cite{MON06}.

We use the spatially extended FitzHugh-Nagumo (FHN) system \cite{Fit61,Nag62}, which has a
cubic nonlinearity, as a generic model of SD waves
\begin{eqnarray}
\label{eq:fhnPlusNLC-1}
  \frac{\partial u}{\partial t} &=& u - \frac{1}{3} u^3 - v
  + D  \frac{\partial^2 u}{\partial x^2} \nonumber \\
  &&+ \mbox{nonlocal coupling} \\
\label{eq:fhnPlusNLC-2}
  \frac{\partial v}{\partial t} &=& \epsilon ( u + \beta - \gamma v)\nonumber \\
  &&+ \mbox{nonlocal coupling}\,\,. 
\end{eqnarray}
The model approximates the cerebral cortex as a two-dimensional
surface with the ability to support sustained SD wave propagation.  As
a generic model this system does not make an explicit distinction
between the various species involved in SD.  In effect it lumps
together sodium inward currents and extracellular potassium
concentration $[K^+]_o$ into a single activator variable $u$ and their
combined kinetics into the cubic source term.  Likewise, a single
inhibitor variable $v$ is related to recovery processes, such as
effective regulation of $[K^+]_o$ by $Na^+$-$K^+$ ion pumps and the
glia-endothelial system \cite{HER94,Kag00,Som01}.  Whether a
transcellular or extracellular route is taken, is at this level not
specified.  The main reason we use the FHN mechanism as the
reaction-diffusion part of the model is that it has been shown to
successfully reproduce the two-dimensional spatio-temporal pattern of
SD \cite{DAH04b,DAH97}.  Our study is essentially based on these
pattern formation properties of SD waves and less on its detailed
biophysical mechanism.

We extend the FHN system to encompass cortical lateral interactions, i.\,e.,
connections running parallel to the cortical surface. They are
accounted for in the form of nonlocal coupling terms
\begin{eqnarray}
\label{eq:nlc}
K\left[s(x+\delta)-2s(x)+s(x-\delta)\right].
\end{eqnarray}
The signal $s$ can either be the activator $u$ or inhibitor $v$.  A
connection in the cortex can extend over several millimeters and it
either mediates competitive or cooperative interactions.  The parameter
$\delta$ describes the connection length and  the coupling strength  $K$ 
of  the interaction.  

Lateral connections in the cortex can form clusters at regular
intervals \cite{GIL89}. Their structure and how they might interact
with SD waves will be further considered in Sec.~\ref{sec:conPat}
Until then, we consider only one nonlocal coupling term occurring
either in the activator (\ref{eq:fhnPlusNLC-1}) or inhibitor
(\ref{eq:fhnPlusNLC-2}) balance equation, with fixed values for
$\delta$ and $K$. This leads to four different coupling schemes: two
of cross coupling (CC) activator and inhibitor, and two in which each
dynamic variable is coupled via Eq.~(\ref{eq:nlc}) into its own balance equation (NCC).

\section{Suppression of waves by nonlocal interaction}

\begin{figure}[!b]
\centerline{\includegraphics[width=0.5\textwidth]{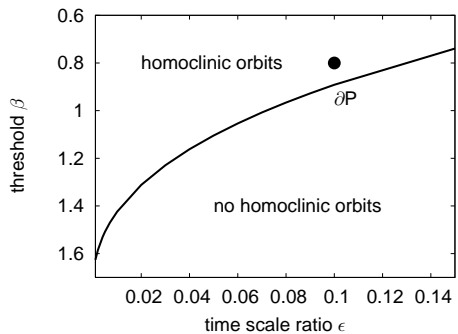}}
\caption{\label{fig:pb} Parameter space of the FHN system at the
section $\gamma=0.5$. $\partial P$ is the propagation boundary.  Below
$\partial P$ any confined perturbation of arbitrary profile decays
eventually.  Above $\partial P$ some wave profiles are stable and
propagate with constant speed. They correspond to homoclinic orbits in
a co-moving frame. The simulations have been done with a FHN system at
$\beta=0.8$ and $\epsilon=0.1$ (solid black circle). A successful
suppression of reaction-diffusion waves by nonlocal coupling indicates
a shift of $\partial P$ of the combined system beyond the point at
$\beta=0.8$ and $\epsilon=0.1$.}
\end{figure}

The investigations were done in a one-dimensional spatial FHN system.
We chose the parameter of the FHN system such that it is without the
nonlocal coupling term in
Eqs.~(\ref{eq:fhnPlusNLC-1}-\ref{eq:fhnPlusNLC-2}) above but near to
the excitability boundary of a one-dimensional system. This boundary
was obtained by transforming the system into a co-moving frame and
searching for homoclinic orbits. These orbits correspond to pulse
solutions in the original coordinates.  The excitability boundary
($\partial P$) is defined by a saddle-node bifurcation at which the
stable and unstable homoclinic orbit disappear \cite{KUZ95}. The
parameter values of $\beta$, $\epsilon$, and $\gamma$ at which
homoclinic orbits disappear constitute a boundary  of
codimension one. In other words, $\partial P$ separates the parameter
plane into a regime where local stimulations is transmitted without
damping and a regime where such sustained 1D reaction-diffusion waves
do not exist (Fig.~\ref{fig:pb}).

We chose a FHN system near $\partial P$ because the transient nature
of the observed symptomatic and electrophysiological events during
migraine suggest such a regime \cite{LAS41,HAD01}.  In the regime
below but close to $\partial P$ transient wave forms exist in a
one-dimensional system \cite{DAH07a}. Above $\partial P$ transient
wave forms exist in two-dimensional systems untill excitability
reaches a boundary called $\partial R$. There, sustained wave
segments, called ``critical fingers'', propagate without reentering
tissue (c.f. Fig.~\ref{fig:ret}).  The regime between $\partial P$ and
$\partial R$ is therefore called sub-excitable.  The regime in which
this transition takes place is also well investigated in chemical
model systems in experiment and theory, for a review see
\cite{Mik06}.

\begin{figure}[!b]
\centerline{\includegraphics[width=0.75\textwidth]{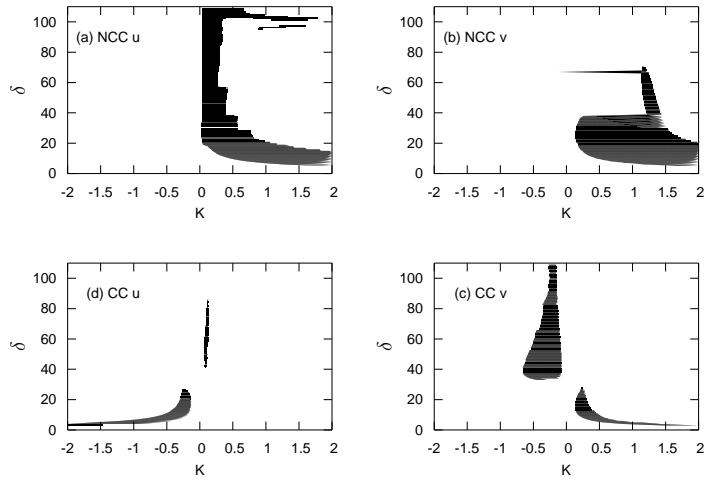}}
\caption{\label{fig:controPlane} Parameter plane ($K,\delta$) of the
nonlocal control term. Black areas indicate successful suppression of
wave propagation. ({\sf a}) Non-cross coupling (NCC) in the activator
equation (\ref{eq:fhnPlusNLC-1}). ({\sf b}) NCC in the inhibitor
equation (\ref{eq:fhnPlusNLC-2}). ({\sf c}) Cross coupling (CC) in the
activator equation (\ref{eq:fhnPlusNLC-1}). ({\sf d}) CC in the
inhibitor equation (\ref{eq:fhnPlusNLC-2}).  }
\end{figure}

To investigate the influence of various nonlocal connectivity schemes
on wave propagation in the regime of sub-excitability, we start by
setting a super-threshold stimulation in the one-dimensional system,
choosing a particular FHN system with parameter values $\beta=0.8$,
$\epsilon=0.1$, $\gamma=0.5$, and $D=1$.  Once a stable
one-dimensional wave profile is obtained, the nonlocal lateral network
is switched on.  Different networks for various parameter values $K$
and $\delta$ are classified by their effect on the wave. We
distinguish two cases.  Either the wave is suppressed. This indicates
that the excitability boundary $\partial P$ of the combined system is
shifted to higher excitability values (upwards in Fig.~\ref{fig:pb})
into a regime where without the nonlocal coupling pulse solutions
would exit. Or the wave continues to spread, though its profile and
speed might change. From a clinical point of view, the wave
suppression is a desirable control goal for the network achieved
within the solid black regions in the $(K,\delta)$-planes in
Fig.~\ref{fig:controPlane}.

We find that wave propagation can be suppressed with a NCC
(non-cross-coupled) setup only with positive coupling strength $K$.
When the NCC term appears in the activator balance equation, the
desired control goal is achieved largely independent of the
connection length $\delta$ (Fig. \ref{fig:controPlane} a), as long as $\delta$ is in the range of the
wave width, including its refractory tail.   When the nonlocal coupling term appears in the
inhibitor balance equation, a similar picture arises, though waves are 
 suppressed for connection lengths ranging into the refractory
tail of the wave ($\delta>40$) only for a narrow regime of $K$.  Suppression completely fails for  $\delta>70$ (Fig. \ref{fig:controPlane} b).  

\begin{figure}[!tpb]
\centerline{\includegraphics[width=0.5\textwidth]{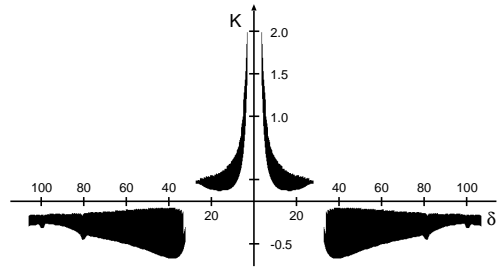}}
\caption{\label{fig:mex} The "Mexican hat" connectivity profile of
successful wave suppression for cross coupling (CC) is clearly visible
when the control plane is rotated and the space coordinate $\delta$ is
plotted as the distance ranging from negative to positive
values. Shown is the successful control area (black) for the CC term
appearing in the inhibitor equation (\ref{eq:fhnPlusNLC-2}).  When the
CC term is in the activator equation (\ref{eq:fhnPlusNLC-1})
the profile of the Mexican-hat connectivity  is inverted.  }
\end{figure}

Cross coupling of inhibitor and activator achieves the desired control
goal for both positive and negative coupling strengths $K$, depending
on the connection length $\delta$ (Fig. \ref{fig:controPlane}
c-d). The area in the parameter plane $(\delta,K)$ where this control
goal is achieved resembles a Mexican-hat-type network
connectivity. This is readily seen in Fig.~\ref{fig:mex}. When the
nonlocal term appears in the inhibitor balance equation
(\ref{eq:fhnPlusNLC-2}) the regimes of successful control in the $K$
direction is much wider (Fig. \ref{fig:controPlane} d) than the regime
for cross coupling in the activator balance equation
(\ref{eq:fhnPlusNLC-1}, Fig. \ref{fig:controPlane} c).

\section{On the nature of nonlocal coupling in migraine}
\label{sec:conPat}

In the previous section, we have shown which network failures lead to
the emergence of reaction-diffusion waves. From this alone it is not
deducible whether some of these network failures can also explain
the anomalous cortical state  in the interval between migraine attacks.  It
would be a plausible hypothesis, however, that the same network
changes that cause the ictal migraine events, i.\,e., SD waves, lead
to the anomalous interictal state.  Changes causing the latter have
been attributed to abnormal aspects in early visual processing in the
cortex \cite{SHE01}.  Various, seemingly contradictory explanations
have been given, such as lack of both intra-cortical inhibition and
excitation.  They are also referred to as cortical hypo- or
hyperexcitability (see \cite{DAH04a} and references therein).

There is substantial work on the functional role of lateral
connectivity for cortical processing, but little is known how the
mechanism of SD is coupled to it. Evidence supporting a coupling comes
from two independent sources. On the one hand, there is the structure
of the hallucinatory aura patterns, in particular the typical zigzags
(see Fig.~\ref{fig:ret}). Such patterns were suggested to reflect the
cortical network organization \cite{SCH80,DAH00a}.  The main idea is
that the approaching wave initially affects cortical cells which
possess the highest spontaneous activity and are clustered in
patches. Within these patches the neuronal response properties remain
relatively constant. Their feature distribution corresponds to the
organization of the receptive field structure in the cortex. For
example, the connection pattern in the visual cortex is unspecific in
the immediate vicinity of each neuron, while long-range connections
primarily run between so-called iso-orientation columns \cite{GIL89}.
Cells in an orientation column have the same oriented receptive field
structure, thus they are responsive to edges with the same
orientation.  These edges are literally seen during a migraine attack
as the building blocks of the hallucinatory zigzag pattern. Therefore,
it is reasonable to assume that the SD wave interacts with this neural
network structure in form of a synchronization process that occurs at
the front of the SD wave and extends over the typical spatial length
scale of iso-orientation columns.

The other line of evidence comes from {\it in vivo} studies in animal
research on SD.  In \cite{HER94} Herreras et al. showed that a
synchronization of the firing pattern is possible up to the order of
millimeters ahead of the SD wave.  The peculiarity of this activity is
that it is resistant to synaptic transmission blockade. This led to
the hypothesis of direct neuron-to-neuron communication by previously
closed gap junctions. They suggested that SD propagates through
transcellular pathways using a reaction-diffusion mechanism. Computer
simulations of Shapiro support this scheme \cite{SHA01}.  A complete
description of SD, however, must additionally include the full network
connectivity of synaptic transmission when SD occures close to a
bifurcation.  Such a description of SD is beyond the scope of the
present study.  The time and space scales of these dynamics differ by
several orders of magnitude such that a separate treatment is
justified.  Therefore we investigate the stability of the suggested
neuron-to-neuron communication by gap junctions separately.  We assume
that the way this transcellular pathway interacts with synaptic
transmission is in principle described in the previous section.

\section{Time-delayed diffusive electrical coupling}

In a previous study \cite{HAU06} we used the FHN system modeling two
individual neurons with a diffusive coupling in the activator
variable.  We showed that two FHN-neurons, each oscillating under its
own source of noise, can synchronize.  The application of time-delayed
feedback to only one of two subsystems was shown to change coherence
and time scales globally. Time delayed feedback is also able to induce
stochastic synchronization under certain conditions.  This motivates
the approach pursued here to examine a time-delayed coupling between
two identical neurons. Since the time-delay can introduce rich
dynamics we study the case without random fluctuations.

\begin{figure}
\centerline{\includegraphics[angle=-90,width=\textwidth]{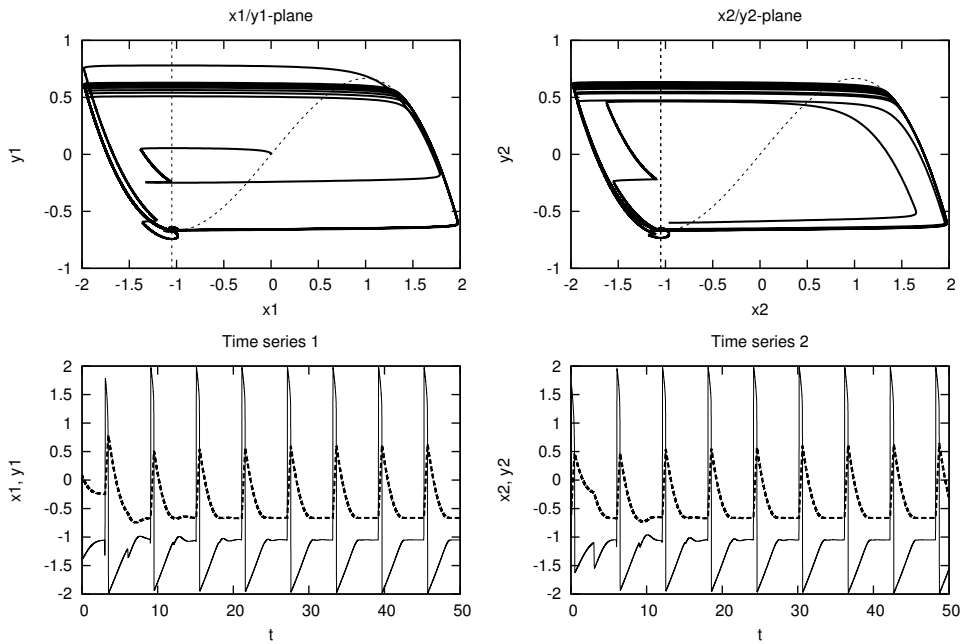}}
\caption{\label{fig:phaseportrait}Simulation for two symmetrically
  coupled identical FHN-subsystems.  Parameter values are
  $\epsilon=0.01, a=1.05, \tau=3,  C=0.4$.  Shown are phase space sections
  corresponding to the individual subsystems and their associated time
  series.  After a short transient effect  the
  combined system settles into a stable firing oscillation.  The
  initial history functions used for this simulation are solutions to
  the uncoupled subsystems.}
\end{figure}

To distinguish this system of two neurons from the spatially extended
FHN reaction diffusion system describing the cortical tissue in
Eqs.~(\ref{eq:fhnPlusNLC-1}-\ref{eq:fhnPlusNLC-2}), we use the
variables $x$ and $y$ with a subscript 1 and 2 identifying the two
neurons. The variables represent the membrane potential and the
gating, respectively. The diffusive coupling occurs in the membrane
potential. This is a discrete model of a gap junction-mediated
electrical coupling, because ionic currents through gap-junctions give
rise to strong electrical coupling of the neurons. Gating mechanisms
of neuronal gap junctions have not been described as yet. Therefore,
we do not consider any gating.  But we include a time delay $\tau$
because if the spread in the transcellular pathway is diffusion
limited, as the slow propagation speed of SD clearly suggests, the
transmission time can be in the order of the excitation cycle.
\begin{eqnarray}
\label{eq:gapJunction}
  \frac{\partial x_1}{\partial t} &=& x_1 - \frac{1}{3} x_1^3 - y_1 + C(x_2(t-\tau)-x_1)
\nonumber \\
  \frac{\partial y_1}{\partial t} &=& \epsilon ( x_1 + a )\nonumber \\ 
  \frac{\partial x_2}{\partial t} &=& x_2 - \frac{1}{3} x_2^3 - y_2 + C(x_1(t-\tau)-x_2)
\nonumber \\
  \frac{\partial y_2}{\partial t} &=& \epsilon ( x_2 + a )\nonumber 
\end{eqnarray}
Individual neurons have only one stable fixed point (for $a>1$). It is
readily shown that when $\tau=0$, the coupled system also has only one
stable fixed point.  With a non-vanishing delay time the phase space
is infinite dimensional.  Then, the fixed point is given by the four
coordinates above as well as their respective history functions of length
$\tau$, which need to be constant. Along the lines of \cite{JAN03}, it
can be shown that this fixed point is stable.  We find that for
adequate parameter values $C$, $\epsilon$ and $\tau$, the system is
multi-stable. It can avoid the stable fixed point and instead
exhibit a mutual resonance phenomenon.  In the 4D phase space section
at time $t$, this results in a stable firing oscillation of period $2
\tau$ between the two sub-systems (see Fig.\ref{fig:phaseportrait}).
Thus, for two FHN-Neurons in the excitable regime, a non-vanishing
delay enables a synchronous operation of the two subsystems.

\section{Discussion}

We showed that certain control schemes of an inhibitor-activator type
system shift the emergence of wave propagation towards higher values
of excitability. The control we investigated is of the form of a
nonlocal coupling given in Eq.~(\ref{eq:nlc}). This nonlocal
transaction was added to the reaction-diffusion mechansim either in
the inhibitor or the activator balance equation.  The sum of all
individual cross coupling terms that achieve a clinically desirable
control goal takes the shape of an upright or inverted Mexican hat,
respectively.  This supports our assumption that the nonlocal coupling
results from intrinsic lateral cortical connections.

Dichotomic lateral interaction is an architecture widely used in
models of topographic feature maps. The prototypical example of such
maps is the orientation preference in primary visual cortex, which is
activated by the SD wave.  The link between SD and the neuronal
network architecture is still missing. One possibility is that
gap-junction-mediated oscillatory patterns trigger SD. If so, these
oscillatory patterns are likely to be modulated by lateral synaptic
connections, although their existence is in general resistant to
synaptic transmission blockade \cite{HER94}.  However, when SD is
close to the bifurcation of the onset of wave propagation, as
suggested by the spatio-temporal patterns (e.g. in Fig.~\ref{fig:ret}),
therapy might target network connectivity as to prevent spread.

Although we are still far from modeling the full mechanism of migraine
with aura, neural network models have become sophisticated enough to
constrain and validate possible underlying cortical circuitry of
involved subsystems.  To understand the origin of the
gap-junction-mediated oscillatory patterns better, we performed
simulations in a system of two gap-junction-coupled neurons. We showed
that a time-delay is sufficient to produce sustained oscillations in
an otherwise merely excitable ensemble. Thus, opening gap junctions
between neurons, which are closed in a healthy state, can explain
a localized pathological synchrony in the cortex when there exits a time delay.

To summarize, in modeling migraine a major objective is to
understanding cortical susceptibility to focal neurological symptoms
in terms of neural circuitry \cite{DAH07a,MON06,REG94}. This could
open up to us new strategies for therapy using methods of controlling
complex dynamics.  Control of complex dynamics has evolved during the
last decade as one of the central issues in applied nonlinear science
\cite{SCH07}.  Progress toward clinical implementation of nonlinear
methods has been done so far in neurology in particular in Parkinson's
disease, a neurological diseases also characterized by pathological
brain synchrony. There, techniques based on control of complex
dynamics \cite{TAS03} are now tested in clinical studies and
fundamentally novel therapy methods are being evolved \cite{HAU07}.
It is hoped that this success can be expanded.

\section*{Acknowledgments}

This work has been partially supported by the DFG within the framework
of Sfb 555 and the Sachbeihilfe DA-602/1-1.

%
%


\end{document}